\documentclass[prl,floatfix, twocolumn,showpacs]{revtex4}%
\usepackage{amsmath}
\usepackage{graphicx}
\usepackage{amsfonts}
\usepackage{amssymb}%
\begin{document}
\title{Quantum non-local effects with Bose-Einstein condensates}
\author{F. Lalo\"{e} $^{a}$ and W. J. Mullin $^{b}$}
\affiliation{$^{a}$Laboratoire Kastler Brossel, ENS, UPMC, CNRS; 24 rue Lhomond, 75005 Paris, France\\
$^{b}$Department of Physics, University of Massachusetts, Amherst,
Massachusetts 01003 USA}

\begin{abstract}
We study theoretically the properties of two Bose-Einstein condensates in different spin
states, represented by a double Fock state.\ Individual
measurements of the spins of the particles are performed in transverse
directions, giving access to the
relative phase of the condensates. Initially, this phase is completely
undefined, and the first measurements provide random results.\ But a fixed
value of this phase rapidly emerges under the effect of the successive quantum
measurements, giving rise to a quasi-classical situation where all spins have
parallel transverse orientations.\ If the number of measurements reaches its maximum (the number of particles), quantum effects show up again,
giving rise to violations of Bell type inequalities. The violation
of BCHSH inequalities with an arbitrarily large number of spins may be comparable (or even equal) to that obtained with
two spins.

\end{abstract}

\maketitle

The notion of non-locality in quantum mechanics (QM) takes its roots in a
chain of two theorems, the EPR (Einstein Podolsky Rosen) theorem \cite{EPR}
and its logical continuation, the Bell theorem.\ The EPR\ theorem starts from
three assumptions (Einstein realism, locality, the predictions of quantum
mechanics concerning some perfect correlations are correct) and proves that QM
is incomplete: additional quantities, traditionally named $\lambda$, are
necessary to complete the description of physical reality.\ The Bell theorem
\cite{Bell,Bell-book} then proves that, if $\lambda$ exists, the predictions of QM concerning other imperfect correlations cannot always be correct.\ The
ensemble of the three assumptions: Einstein realism, locality, all predictions
of QM are correct, is therefore self-contradictory; if Einstein realism is
valid, QM is non-local.\ Bohr \cite{Bohr} rejected Einstein realism because, in his view, the notion of physical reality could not correctly be applied to microscopic quantum systems, defined independently of the measurement apparatuses. Indeed, since EPR consider a system of two microscopic particles, which can be \textquotedblleft seen\textquotedblright\ only with the help of measurement apparatuses, the notion of their independent physical reality is open to discussion.

Nevertheless, it has been pointed out recently \cite{FL, MKL} that the
EPR\ theorem also applies to macroscopic systems, namely Bose-Einstein (BE)
condensates in two different internal states. The $\lambda$ introduced by
EPR\ then corresponds to the relative phase of the condensates, i.e. to
macroscopic transverse spin orientations, physical quantities at a human
scale; it then seems more difficult to deny the existence of their reality,
even in the absence of measurement devices. This gives even more strength to
the EPR argument and weakens Bohr's refutation.\ It is then natural to ask whether the Bell theorem can be transposed to this stronger case.

The purpose of this article is to show that it can.\ We consider an ensemble of
$N_{+}$ particles in a state defined by an orbital state $u$ and a spin state
$+$, and $N_{-}$ particles in the same state with spin orientation $-$. The
whole system is described quantum mechanically by a double Fock state, that
is, a \textquotedblleft double BE\ condensate\textquotedblright: 
\begin{equation}
\mid\Phi>~=~\left[  \left(  a_{u,+}\right)  ^{\dagger}\right]  ^{N_{+}}\left[
\left(  a_{u,-}\right)  ^{\dagger}\right]  ^{N_{-}}\mid\text{vac}.> \label{1}%
\end{equation}
where $a_{u,+}$ and $a_{u,-}$ are the destruction operators associated with the two populated single-particle states and $\mid$vac$.>$ is the vacuum state. We
introduce a sequence of transverse spin measurements that leads to quantum
predictions violating the so called BCHSH\ \cite{BCHSH, CS}
Bell inequality. This is reminiscent of the work of Mermin \cite{Mermin-1},
who finds exponential violations of local realist inequalities with $N$-particle
spin states that are maximally entangled. By contrast, here we consider the simplest way in which many bosons can be put in two different internal levels, with a $N$-particle state containing only the minimal possible correlations, those due to statistics. We find violations of inequalities that are the same order of magnitude as with the usual singlet spin state and may actually saturate the Cirel'son bound \cite{Cirel}.

Double Fock states are experimentally more accessible and much less sensitive to dissipation and decoherence than maximally entangled states \cite{DBB}. Considering a system in a double Fock state, we assume that a series of rapid spin measurements can be performed and described by the usual QM postulate of measurement, without worrying about decoherence between the measurements, thermal effects, etc.

The operators associated with the local density of particles and spins can be
expressed as functions of the two fields operators $\Psi_{\pm}(\mathbf{r})$
associated with the two internal states $\pm$ as: $n(\mathbf{r})=~~\Psi
_{+}^{\dagger}(\mathbf{r})\Psi_{+}(\mathbf{r})+\Psi_{-}^{\dagger}%
(\mathbf{r})\Psi_{-}(\mathbf{r})$, $\sigma_{z}(\mathbf{r})=~\Psi_{+}^{\dagger
}(\mathbf{r})\Psi_{+}(\mathbf{r})-\Psi_{-}^{\dagger}(\mathbf{r})\Psi
_{-}(\mathbf{r})$, while the spin component in the direction of plane $xOy$
making an angle $\varphi$ with $Ox$ is: $\sigma_{\varphi}(\mathbf{r}%
)=e^{-i\varphi}\Psi_{+}^{\dagger}(\mathbf{r})\Psi_{-}(\mathbf{r}%
)+~e^{i\varphi}\Psi_{-}^{\dagger}(\mathbf{r})\Psi_{+}(\mathbf{r})$.
Consider now a measurement of this component performed at point $\mathbf{r}$
and providing result $\eta=\pm1$. The corresponding projector is:
\begin{equation}
P_{\eta=\pm1}(\mathbf{r,\varphi})=\frac{1}{2}\,\left[  n(\mathbf{r}%
)+\eta~\sigma_{\varphi}(\mathbf{r})\right]  \label{4}%
\end{equation}
and, because the measurements are supposed to be performed at different points
(ensuring that these projectors all commute) the probability $\mathcal{P}%
(\eta_{1},\eta_{2},...\eta_{N})$ for a series of results $\eta_{i}\pm1$ for spin measurements
at points $\mathbf{r}_{i}$ along directions
$\varphi_{i}$ can be written as:
\begin{equation}
~<\Phi\mid P_{\eta_{1}}(\mathbf{r}_{1}\mathbf{,\varphi}_{1})\times P_{\eta
_{2}}(\mathbf{r}_{2}\mathbf{,\varphi}_{2})\times....P_{\eta_{N}}%
(\mathbf{r}_{N}\mathbf{,\varphi}_{N})\mid\Phi> \label{5}%
\end{equation}

We now substitute the expression of $\sigma_{\varphi}(\mathbf{r})$ into
(\ref{4}) and (\ref{5}), exactly as in the calculation of ref. \cite{FL}, but
with one difference: here we do not assume that the number of measurements is
much smaller than $N_{\pm}$, but equal to its maximum
value $N=N_{+}+N_{-}$. In the product of projectors appearing in
(\ref{5}), because all $\mathbf{r}$'s are different, commutation allows us to
push all the field operators to the right, all their conjugates to the left;
one can then easily see that each $\Psi_{\pm}(\mathbf{r})$ acting on
$\mid\Phi>$ can be replaced by $u(\mathbf{r})\times a_{u,\pm}\,$, and
similarly for the Hermitian conjugates. With our initial state, a non-zero
result can be obtained only if exactly $N_{+}$ operators $a_{u,+}$ appear
in the term considered, and $N_{-}$ operators $a_{u,-}$; a similar condition
exists for the Hermitian conjugate operators. To express these conditions, we
introduce two additional variables.\ As in \cite{FL}, the first variable
$\lambda$ ensures an equal number of creation and destruction operators in the
internal states $\pm$ \ through the mathematical identity:
\begin{equation}
\int_{-\pi}^{\pi}\frac{d\lambda}{2\pi}~e^{in\lambda}~=~\delta_{n,0}\label{6}%
\end{equation}
The second variable $\Lambda$ expresses in a similar way that the difference
between the number of destruction operators in states $+$ and $-$ is exactly
$N_{+}-N_{-}$, through the integral:
\begin{equation}
\int_{-\pi}^{\pi}\frac{d\Lambda}{2\pi}~e^{-in\Lambda}~e^{i(N_{+}-N_{-})\Lambda
}=~\delta_{n,N_{+}-N_{-}}\label{7}%
\end{equation}
The introduction of the corresponding exponentials into the product of
projectors (\ref{4}) in (\ref{5}) provides the expression (c.c. means complex
conjugate):
\begin{equation}
\prod\limits_{j=1}^{N}\,\left\vert u(\mathbf{r}_{j})\right\vert ^{2}~\frac
{1}{2}\left[  e^{i\Lambda}+e^{-i\Lambda}+\eta_{j}\left(  e^{i\left(
\lambda-\varphi_{j}+\Lambda\right)  }+\text{c.c.}\right)  \right]  \label{8}%
\end{equation}
where, after integration over $\lambda$ and $\Lambda$, the only surviving
terms are all associated with the same matrix element in state $\mid\Phi>$
(that of the product of $N_{+}$ operators $a_{u,+}^{\dagger}$ and $N_{-}$
operators $a_{u,-}^{\dagger}$ followed by the same sequence of destruction
operators, providing the constant result $N_{+}!N_{-}!$).\ We can thus write
the probability as:
\begin{widetext}%
\begin{equation}
\mathcal{P}(\eta_{1},\eta_{2},...\eta_{N})\sim~\int_{-\pi}^{\pi}\frac
{d\lambda}{2\pi}\int_{-\pi}^{+\pi}\frac{d\Lambda}{2\pi}~e^{i(N_{+}-N_{-})\Lambda
}\prod\limits_{j=1}^{N}\left\{  \left\vert u(\mathbf{r}_{j})\right\vert
^{2}\frac{1}{2}\left[  e^{i\Lambda}+e^{-i\Lambda}+\eta_{j}\left(  e^{i\left(
\lambda-\varphi_{j}+\Lambda\right)  }+\text{c.c.}\right)  \right]\right\}
\label{9}%
\end{equation}
or, by using $\Lambda$ parity and changing one integration variable
($\lambda^{\prime}=\lambda+\Lambda$), as:
\begin{equation}
\mathcal{P}(\eta_{1},\eta_{2},...\eta_{N})\ =~\frac{1}{2^{N}C_{N}}~\int_{-\pi
}^{+\pi}\frac{d\Lambda}{2\pi}~\cos\left[  (N_{+}-N_{-})\Lambda\right]
\int_{-\pi}^{+\pi}\frac{d\lambda^{\prime}}{2\pi}\prod\limits_{j=1}%
^{N}\left\{  \cos\left(  \Lambda\right)  +\eta_{j}\cos\left(  \lambda
^{\prime}-\varphi_{j}\right)  \right\}  \label{10}%
\end{equation}
%\end{widetext}
The normalization coefficient $C_{N}$ is readily obtained by writing that the
sum of probabilities of all possible sequences of $\eta$'s is 1 (this step
requires discussion; we come back to this point at the end of this article):
\begin{equation}
C_{N}=\int_{-\pi}^{+\pi}\frac{d\Lambda}{2\pi}~\cos\left[  (N_{+}-N_{-}%
)\Lambda\right]  ~\left[  \cos\left(  \Lambda\right)  \right]  ^{N}\label{11}%
\end{equation}
Finally, we generalize (\ref{10}) to any number of measurements
$M<N$.\ A sequence of $M$
measurements can always be completed by additional $N-M$ measurements, leading to 
probability (\ref{10}). We can therefore take the sum of (\ref{10}) over all possible
results of the additional $N-M$ measurements to obtain the probability for any $M$ as:
%\begin{widetext}%
\begin{equation}
\mathcal{P}(\eta_{1},\eta_{2},...\eta_{M})=
\frac{1}{2^{M}C_{N}}~\int_{-\pi}^{+\pi}\frac{d\Lambda}{2\pi}~\cos\left[
(N_{+}-N_{-})\Lambda\right]  \left[  \cos\Lambda\right]  ^{N-M}\int_{-\pi
}^{+\pi}\frac{d\lambda^{^{\prime}}}{2\pi}\prod\limits_{j=1}^{M}\left\{
\cos\left(  \Lambda\right)  +\eta_{j}\cos\left(  \lambda^{\prime}%
-\varphi_{j}\right)  \right\}  \label{12}%
\end{equation}
The $\Lambda$ integral can be replaced by twice the integral between $\pm
\pi/2$ (a change of $\Lambda$ into $\pi-\Lambda$ multiplies the function by
$(-1)^{N_{+}-N_{-}+N-M+M}=1$).
If $M\ll N$, the large power of $\cos\Lambda$ in the first integral
concentrates its contribution around $\Lambda\simeq0$, so that a good
approximation is $\Lambda =0$.\ We then recover the results of refs
\cite{FL, MKL}, with a single integral over $\lambda$ defining the relative
phase of the condensates (Anderson phase), initially completely undetermined, so that the first spin measurement provides a completely random result. But the phase rapidly emerges under the effect
of a few measurements, and remains constant \cite{Juha, CD, more}; it takes a different value for each realization of the experiment, as
if it was revealing the pre-existing value of a classical quantity.\ Moreover, when $\cos\Lambda$ \ is
replaced by 1, each factor of the product over $j$ remains positive (or zero), leading to a result similar to that of stochastic local realist theories; the Bell inequalities can then be obtained.\
However, when $N-M$ is small or even vanishes, $\cos\Lambda$ can
take values that are smaller than 1 and the factors may become negative, opening the possibility of
violations.\ In a sense, the additional variable $\Lambda$ controls the amount
of quantum effects in the series of measurements.

We now discuss when these standard QM predictions violate Bell
inequalities.\ We need the value of the quantum average of the product of
results, that is the sum of $\eta_{1},\eta_{2},...\eta_{M}\times\mathcal{P}(\eta_{1},\eta_{2},...\eta_{M})$ over all
possible values of the $\eta$'s, which according to (\ref{12}) is given by:
\begin{equation}
E(\varphi_{1},\varphi_{2},..\varphi_{M})~=\frac{1}{C_{N}}~\int_{-\pi}^{+\pi
}\frac{d\Lambda}{2\pi}\cos\left[  (N_{+}-N_{-})\Lambda\right]  \left[
\cos\Lambda\right]  ^{N-M}\int_{-\pi}^{+\pi}\frac{d\lambda^{\prime}}{2\pi
}\prod\limits_{j=1}^{M}  \cos\left(  \lambda^{^{\prime}}-\varphi
_{j}\right)   \label{13}%
\end{equation}
\end{widetext}
Consider a thought experiment where two condensates in different
spin states (two eigenstates of the $Oz$ spin component) overlap
in two remote regions of space $\mathcal{A}$
and $\mathcal{B\,}$, with two experimentalists Alice and Bob; they measure the
spins of the particles in arbitrary transverse directions (perpendicular to $Oz$) at points of space where the orbital wave functions of
the two condensates are equal.\ All measurements performed by Alice
are made along a single direction $\varphi_{a}$, which plays here the usual
role of the \textquotedblleft setting\textquotedblright\ $a$, while all those
performed by Bob are made along angle $\varphi_{b}$. We assume that Alice retains just the product $A$ of all her measurements, while Bob retains only the
product $B$ of his; $A$ and $B$ are both $\pm1$.

We now assume two possible orientations $\varphi_{a}$ and $\varphi
_{a}^{\prime}$ for Alice, two possible orientations $\varphi_{b}$ and
$\varphi_{b}^{\prime}$ for Bob.\ Within deterministic local realism, for
each realization of the experiment, it is possible to define two numbers
$A$, $A^{\prime}$, both equal to $\pm1$, associated with the two possible
products of results $\eta$ that Alice will observe, depending of her choice of
orientation; the same is obviously true for Bob, introducing $B$ and
$B^{\prime}$.\ Within stochastic local realism \cite{CS, FL-MQ}, $A$ and
$A^{\prime}$ are the difference of probabilities associated with Alice observing $+1$ or $-1$,
i.e. numbers that have values between $+1$ and $-1$. In both cases, the following
inequalities (BCHSH) are obeyed:
\begin{equation}
-2\leq AB+AB^{\prime}\pm(A^{\prime}B-A^{\prime}B^{\prime})\leq2
\label{14}%
\end{equation}

In standard quantum mechanics, of course, \textquotedblleft unperformed
experiments have no results\textquotedblright\ \cite{AP}, and several of the
numbers appearing in (\ref{14}) are undefined; only two of them can be defined after the
experiment has been performed with a given choice of the orientations. Thus,
while one can calculate from (\ref{13}) the quantum average value $<Q>$ of the sum of
products of results appearing in (\ref{14}), there is no special
reason why $<Q>$ should be limited between $+2$ and
$-2$.\ Situations where the limit is exceeded are called \textquotedblleft
quantum non-local\textquotedblright.

We have seen that the most interesting situations occur when the cosines do
not introduce their peaking effect around $\Lambda=0$, i.e. when
$N_{+}=N_{-}$ and $M$ has its maximum value $N$.
Then, for a given $N$, the only remaining choice is how the number of
measurements is shared between $N_{a}$ measurements for Alice and $N_{b}$ for Bob.

Assume first that $N_{a}=1$ (Alice makes one measurement) and therefore
$N_{b}=N-1$ (Bob makes all the others).\ Since we assume that $N_{+}=N_{-}$
and $M=N$, the $\Lambda$ integral in (\ref{13}) disappears, and
the $\lambda$ integral contains only the product of $\cos\left(
\lambda^{\prime}-\varphi_{a}\right)  $ by the $(N-1)$th power of $\cos\left(
\lambda^{\prime}-\varphi_{b}\right)  $, which is straightforward and
provides $\cos\left(  \varphi_{a}-\varphi_{b}\right)  $ times the
normalization integral $C_{N}$.\ The quantum average associated with the product
$AB$ is thus merely equal to $\cos\left(  \varphi_{a}-\varphi_{b}\right)  $,
exactly as the usual case of two spins in a singlet state.\ Then it is
well-known that, when the angles form a \textquotedblleft
fan\textquotedblright\ \cite{fan}\ spaced by $\chi=\pi/4$, a strong violation of
(\ref{14}) occurs, by a factor $\sqrt{2}$, saturating the Cirel'son bound
\cite{Cirel}. A similar calculation can be performed when Alice makes 2
measurements and Bob $N-2$, and shows that the quantum average is now equal to
$\frac{1}{2}\left[  1+\frac{1}{N-1}+(1-\frac{1}{N-1})\cos2\left(  \varphi
_{a}-\varphi_{b}\right)  \right]  ,$ no longer independent of $N.$ If $N=4$,
the maximum of $<Q>$ is $2.28<2\sqrt{2}$, and \emph{rises} to 2.41 as
$N \rightarrow\infty.$ An 
expression for the generalization of the quantum average to any number $P$ and
$N-P$ of measurements by Alice and Bob, respectively, is (with
$\chi=\varphi_{a}-\varphi_{b}$):
\begin{equation}
E(\chi)=\frac{\frac{N}{2}!}{N!}\sum_{k=0}^{\{P/2\}}\frac{P!(N-2k)!}%
{k!(P-2k)!(\frac{N}{2}-k)!}\sin^{2k}\chi\cos^{P-2k}\chi\label{15}%
\end{equation}
where $\{P/2\}$ is the integer part of $P/2$. The maximum of $<Q>$ 
can then be found using a numerical Mathematica routine. Results are
shown for several values of $P$ in Fig. 1. The angles maximizing the quantum
Bell quantity always occur in the fan shape, although the basic angle $\chi$ changes
with $P$ and $N.$ All of the curves where $P$ is held fixed have a finite $<Q>$
limit with increasing $N$, and the optimum values of the angles approach constants.
\ For the curve $P=N/2$, the limit is 2.32 when $N\rightarrow\infty
$, and the fan opening decreases as 1/$\sqrt{N}.$ \begin{figure}[ptb]

\includegraphics[width=3.50in]{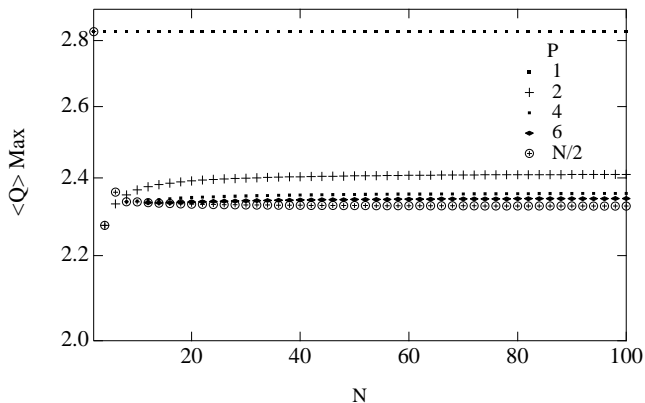}\caption{The maximum of the quantum average $<Q>$ 
for Alice doing $P$ experiments and Bob $N-P$, as a function of
the total number of particles $N$. The usual Bell situation is obtained for $N=2, P=1.$ Local realist theories predict an upper limit of $2$;
large violations of this limit are obtained,
even with macroscopic systems ($N\rightarrow\infty)$.\ If $P=1$, the violation
saturates the Cirel'son limit for any $N$.\ }%
\label{MaxQ}%
\end{figure}

We can also study cases where the number of measurements is $M<N~$: if Bob
makes all his measurements, but ignores one or two of them (independently of
the order of the measurements), when he correlates his results with Alice, the
BCHSH inequality is never violated. All measurements have to be taken into
account to obtain violations. Furthermore, if the number of particles in the
two condensates are not equal, no violation occurs either. Finally, it is
possible to consider cases where we generalize the angles considered: 
experimenter Carole makes measurements at $\varphi_{c}$ and $\varphi_{c}^{\prime}$, and
David at $\varphi_{d}$ and $\varphi_{d}^{\prime}.$ We then find that
a maximization of $<Q>$ reduces to the cases already studied,
where the new angles collapse to the previous angles $\varphi_{a}%
,\cdots,\varphi_{b}^{\prime}$.

For the sake of simplicity, we have not yet discussed some important issues
that underlie our calculations.\ One is related to the so called
\textquotedblleft sample bias loophole\textquotedblright\ (or
\textquotedblleft detection/efficiency loophole\textquotedblright) and to the
normalization condition (\ref{11}), which assumes that one spin is detected at
each point of measurement.\ A more detailed study (see second ref. \cite{FL}) should
include the integration of each $\mathbf{r}$
in a small detection volume and the possibility
that no particle is detected in it.
This is a well-known difficulty, which already
appears in the usual two-photon experiments \cite{CS}, where most photons are
missed by the detectors.\ If this loophole still raises a real experimental
challenge, the difficulty can be resolved in theory 
by assuming the presence of additional spin-independent
detectors \cite{Bell, CS}, which ensure the detection of one particle in each detector and
create appropriate initial conditions (see for instance
\cite{Bell-preliminary-det} for a description of an experiment with
veto detectors).\ We postpone this discussion to another article \cite{LM}. A
second issue deals with the definition of the local realist quantities
$A$, $B$, etc.\ For two condensates, we have a slightly different
situation than in the usual EPR\ situation: the local realist reasoning leads
to the existence of a well-defined phase $\lambda$ between the condensates \cite{FL}, not necessarily to deterministic properties of the individual particles.\ Fortunately, Bell inequalities can also be derived within stochastic local realist theories \cite{Bell-book, CS} (see also for instance \cite{Mermin-1} or appendix I
of \cite{FL-MQ}), and this difference is not a problem \cite{LM}.

In conclusion, strong violations of local realism may
occur for large quantum systems, even if the state is a
simple double Fock state with equal populations; within present experimental techniques, this seems reachable with $N\sim10$ or $20$. We have assumed that the measured quantity is
the product of many microscopic measurements, not their sum,
which would be macroscopic; a product of results remains sensitive to the last
measurement, even after a long sequence of others.\ Curiously, for very few measurements only the results are quantum, for many measurements they can be interpreted in terms of a classical phase, but become again strongly quantum when the maximum
number of measurements is reached, a sort of revival of quantum-ness of the system.

Laboratoire Kastler Brossel is \textquotedblleft UMR 8552 du CNRS, de l'ENS,
et de l'Universit\'{e} Pierre et Marie Curie\textquotedblright.

\end{document}